\documentstyle[aps,12pt,tighten,epsfig]{revtex}
\begin{document}

\draft

\title{
The reaction $\mbox{$NN \to NN \phi$}$} near threshold

\author{ {\sc
A.I. Titov$^{a,b}$,
B. K\"ampfer$^a$,
V.V. Shklyar$^b$}}

\address{
$^a$Forschungszentrum Rossendorf, PF 510119, 01314 Dresden, Germany\\[1mm]
$^b$Bogoliubov Theoretical Laboratory, JINR Dubna,
141980 Dubna, Russia}

\maketitle

\begin{abstract}
The sensitivity of polarization observables in the reaction
$NN \to NN \phi$ slightly above the threshold
is studied with respect to
the details of the one-boson exchange model and a
possible admixture of hidden strangeness in the nucleon.
It is shown that the finite-energy predictions differ strongly
from the threshold predictions.
A measurement of the beam - target asymmetry and the $\phi$ decay
anisotropy can help to disentangle the role of various 
reaction mechanisms.\\[3mm]
{\it PACS number(s):\/}
13.75.-n, 14.20.-c, 21.45.+v
\end{abstract}

\section{Introduction}

The investigation of the reaction ${NN\to NN\phi}$
is interesting for several reasons. First, the elementary
total cross section \cite{Sibirtsev}
is an important input for the calculation of the $\phi$ production in
heavy-ion collisions \cite{Ko_PL}.
In this case one might expect some change
of the $\phi$ width \cite{Weise} due to the coupling to the decay
channel $\phi\to K^+K^-$ and peculiarities according to the
in-medium modification of the kaon properties \cite{Kolomeitsev,Grosse}.
Indeed, such measurements are under way with the $4\pi$ detector FOPI
at SIS in GSI/Darmstadt \cite{Kotte}.
The electromagnetic decay channel $\phi \to e^+ e^-$ will be investigated
with the spectrometer HADES \cite{HADES} also in GSI.
A threshold-near measurement
of the total cross section of the reaction $pp\to pp\phi$ has been performed
at SATURNE \cite{Brenschede} and precision measurements of polarization
observables are envisaged with the ANKE spectrometer
at the cooler synchrotron COSY in J\"ulich \cite{Sapozhnikov}.

Second, the $\phi$ meson is thought to consist mainly of strange
quarks, i.e. $s\bar s$, with a rather small contribution of the light
$u,d$ quarks. According to the OZI rule (cf.\ \cite{Ellis} for references)
the $\phi$ production
should be suppressed if the entrance channel does not possess
a considerable admixture of hidden or open strangeness.
Indeed, the recent experiments on the proton annihilation at rest
(cf.\ \cite{Ellis} for a compilation of data) point to
a large apparent violation of the OZI rule,
which is interpreted \cite{Ellis} as a
hint to an intrinsic $s\bar s$ component in the proton.
However, the data can be
explained by modified meson exchange models \cite{LZL93}
without any strangeness content of nucleon or OZI rule violation.
On the other hand
the analysis of the $\pi N$ sigma term \cite{sigmaterm} suggest that
the proton might contain a strange quark admixture as large as 20\%.
Thus this issue remains rather controversial.

Therefore it is tempting to look for other
observables \cite{Ellis,newexp} that are directly related to the
strangeness content of the nucleon. In Ref.~\cite{Titov1} it is shown
that polarization observables of the
$\phi$ photoproduction are sensitive even to a small strangeness admixture
in the proton. Other investigations concern the polarization
observables in $pp\to pp\phi$ reactions
at the threshold \cite{Ellis,Rekalo}.
It is found \cite{Rekalo} that
spin and parity conservation arguments result in a
precise prediction for the beam - target asymmetry
\begin{equation}
C^{BT}_{zz} = \frac
{d\sigma(S_i=1) - d\sigma(S_i=0)}
{d\sigma(S_i=1) + d\sigma(S_i=0)}
\label{as_thr}
\end{equation}
for the $pp\to pp\phi$ reaction at the threshold,
$ C^{BT}_{zz} = 1$,
where $S_i$ is the total spin in the entrance channel.
It is further
claimed \cite{Rekalo} that the elements of the decay density matrix
$\rho_{r\,r'}$ (for notation see below)
amount
$\rho_{00} = 0, \rho_{\pm 1 \pm 1} = \frac12$.
%
% It is further
%claimed \cite{Rekalo} that the elements of the decay density matrix
%\begin{equation}
%\rho_{r r} =
%\frac{T^r \, T^{r *}}{\sum_r \,\vert T^r \vert^2}
%\label{rho_thr}
%\end{equation}
%($T^r$ is the amplitude with $\phi$ polarization $r$, see below)
%amount
%$\rho_{00} = 0, \rho_{\pm 1 \pm 1} = \frac12$.
%where the quantization axis is directed along the velocity of the
%$\phi$ meson.
%
Real experiments, however, are performed at some finite energy
above the threshold, and therefore
the problem arises whether the threshold predictions are modified.
More generally one should investigate the sensitivity of
polarization observables to the strangeness admixture in the proton.
An analysis of the threshold-near behavior of the 
unpolarized cross section
is a necessary prerequisite, of course.

In the present work we employ a one-boson exchange (OBE) model to
consider these mentioned issues. In Section II we present expressions for the
kinematics, cross sections and spin density matrix. In Section III we
describe our OBE approach. Section IV is devoted to parameter fixing.
The threshold limit is analyzed in Section V.  
The results at finite energy above threshold are discussed in Section VI and
summarized in Section VII.

\section{Kinematics and observables}

The corresponding  cross section of
$\phi$ production in  the reaction $a+b\to c+d+\phi$, where $a,b$ and $c,d$
label the incoming and outgoing nucleons (cf. Fig.~1),
is related
to the invariant amplitude $T$ as
\begin{eqnarray}
d\sigma =
\frac{1}{2(2\pi)^5\sqrt{s(s-4M_N^2)}} \,|T|^2 \,
\frac{d{\bf p}_c}{2E_c} \,
\frac{d{\bf p}_d}{2E_d} \,
\frac{d{\bf q}}{2E_\phi} \,
\delta^{(4)}(P_i - P_f).
\label{sig1}
\end{eqnarray}
where
$p_n=(E _n,\, {\bf p}_n)$ with $n=a,b,c,d$ and $q=(E_\phi, {\bf q})$
are the four-momenta of the nucleons and the 
$\phi$ meson in the center of mass system (c.m.s.), respectively.
Hereafter $\theta$ denotes the polar $\phi$ meson angle, $s=E_a+E_b$
is the total c.m.s. energy, and
$P_{i,f}$ is the total 4-momentum of the initial or final states.
We use a coordinate system with
${\bf z} \parallel {\bf p}_a$,
${\bf y} \parallel {\bf p}_a{\bf \times q}$.
%${\bf x} = {\bf y}{\bf \times z}$.
Among the five independent variables for
describing the final state we choose
$E_\phi,\,\Omega_\phi\,$ and $\Omega_c$. Then the energy $E_c$
of particle $c$ is defined by
\begin{eqnarray}
E_c=\frac{AB\pm C\sqrt{B^2-M_N^2(A^2-C^2)}}{A^2-C^2},
\label{eq1}
\end{eqnarray}
with
%\begin{eqnarray}
$A=2(\sqrt{s}-E_\phi)$,
$B=s-2E_\phi\sqrt{s}+M_\phi^2$, 
$C=2q\cos \theta_{qp_c}$,
%\end{eqnarray}
where $\theta_{qp_c}$ is the angle between ${\bf q}$ and ${\bf p_c}$
determined by
%\begin{eqnarray}
$\cos\theta_{qp_c} =
\cos\theta\cos\theta_c
+\sin\theta\sin\theta_c\cos(\varphi-\varphi_c)$;
%\end{eqnarray}
the unindexed polar and azimuthal angles $\theta$ and $\phi$
belong to the vector ${\bf q}$.
In calculations one has to use both solutions of Eq.~(\ref{eq1})
according to the considered kinematics.
Finally, the fivefold differential cross section reads
\begin{eqnarray}
\frac{d^5\sigma}{dE_\phi d\Omega_\phi d\Omega_c}
=
\frac{1}{8(2\pi)^5\sqrt{s(s-4M_N^2)}} \,
|T|^2 \,
\frac{qp_c^2}{|A p_c+C E_c|}.
\label{dcs}
\end{eqnarray}
The total cross section is found by integration of over
the available phase space.

%The differential beam - target asymmetry is defined as the ratio 
%\begin{equation}
%C^{BT}_{zz}(E_\phi,\Omega_\phi, d\Omega_{p'}) = \frac
%{d^5\sigma(S_i=1) - d^5\sigma(S_i=0)}
%{d^5\sigma(S_i=1) + d^5\sigma(S_i=0)}.
%\label{as_5}
%\end{equation}
One can consider the 
asymmetry Eq.~(1) with integrated cross section or
differential cross sections. In the general case the asymmetry
depends also on $E_\phi$, $\Omega_\phi$ and $\Omega_c$.

We consider the decay angular distribution ${\cal W}$
of the $\phi$ meson
in its helicity system where the $\phi$ is at rest.
The quantization axis ${\bf z'}$ 
%(don't mix it with $\bf z$ axis) 
is chosen 
along the $\phi$ meson velocity in the overall c.m.s.,
and the ${\bf y'}$ axis
is parallel to the vector ${\bf p}_a\times {\bf q}$.
% where ${\bf p}_a$ is beam momentum.
The decay angles $\Theta$, $\Phi$ are defined as polar and azimuthal
angles of the direction of flight of one of the decay particles
in the $\phi$ meson's rest frame.
The decay angular distribution 
%of the vector meson in its rest frame 
then reads
\begin{eqnarray}
\frac{dN}{d\cos\Theta d\Phi}
\equiv 
{\cal W}(\cos\Theta,\Phi)
=\sum_{r,r'} M(r;\Theta,\Phi)\,
\rho_{rr'} \,
M^*(r';\Theta,\Phi),
\end{eqnarray}
where $M(r;\Theta,\Phi)$ is the decay amplitude,
%\begin{eqnarray}
$M(r;\Theta,\Phi)=M_0 \,
d^{1*}_{r,\lambda_\alpha-\lambda_\beta}(\Phi,\Theta,-\Phi)$,
%\label{M1}
%\end{eqnarray}
with $r$ as the helicity of $\phi$ meson;
$\lambda_{\alpha,\beta}$ denotes the helicity of decay particles and 
$|M_0|^2$ is proportional to the decay width;
$D^1_{r, r'}$ stands for Wigner's rotation function.
The angular distribution is normalized as
\begin{eqnarray}
\int {\cal W}(\cos\Theta,\Phi) d\Omega =1.
\end{eqnarray}
The density matrix $\rho_{rr'}$ is expressed in terms of 
the production amplitudes
$T^r_\beta$
\begin{eqnarray}
\rho_{rr'}=\frac{\sum_\beta T^r_\beta \, T^{r'*}_\beta}
{\sum_{r, \beta} T^r_\beta \,T^{r\,*}_\beta},
\label{rhom}
\end{eqnarray}
where $\beta$ denotes a set of unobserved quantum numbers. 

Assuming  that in $\phi\to K^+K^-$ and $\phi\to e^+e^-$ decays
$\lambda_{\alpha}-\lambda_{\beta} = 0$
(i.e. $\lambda_{\alpha}-\lambda_{\beta}= \pm 1$),
and using the explicit form of the rotation function we get 
\begin{eqnarray}
{\cal W}^{K^+K^-}(\cos\Theta,\Phi) & = &
\frac{3}{4\pi}\left[ \right.
\frac12(\rho_{11}+\rho_{-1-1})\sin^2\Theta + \rho_{00}\cos^2\Theta
\nonumber\\
& + &
\frac{1}{\sqrt{2}}(-{\rm Re}\rho_{10}
+{\rm Re}\rho_{-10})\sin 2\Theta\cos\Phi
+\frac{1}{\sqrt{2}}({\rm Im}\rho_{10}
+{\rm Im}\rho_{-10})\sin 2\Theta\sin\Phi\nonumber\\
& - &
{\rm Re}\rho_{1-1}\sin^2\Theta\cos2\Phi
+{\rm Im}\rho_{1-1}\sin^2\Theta\sin2\Phi \left. \right],
\label{KK2}
\end{eqnarray}
\begin{eqnarray}
{\cal W}^{e^+e^-}(\cos\Theta,\Phi) & = &
\frac{3}{8\pi} \left[ \right.
\frac12(\rho_{11}+\rho_{-1-1})(1+\cos^2\Theta) + \rho_{00}\sin^2\Theta
\nonumber\\
& + &
\frac{1}{\sqrt{2}}({\rm Re}\rho_{10}
-{\rm Re}\rho_{-10})\sin 2\Theta\cos\Phi
-\frac{1}{\sqrt{2}}({\rm Im}\rho_{10}
+{\rm Im}\rho_{-10})\sin 2\Theta\sin\Phi\nonumber\\
& + &
\frac12{\rm Re}\rho_{1-1}\sin^2\Theta\cos2\Phi
-\frac12{\rm Im}\rho_{1-1}\sin^2\Theta\sin2\Phi \left. \right].
\label{EE2}
\end{eqnarray}
The decay distributions integrated over the azimuthal angle 
depend only on the diagonal matrix elements,
\begin{eqnarray}
{\cal W}^{K^+K^-}(\cos\Theta) & = &
%\frac{3}{4}(1-\rho_{00})
%\left(1-\frac{1-3\rho_{00}}{1-\rho_{00}}\cos^2\Theta\right)
%\equiv
\frac{3}{4}(1-\rho_{00})
\left(1+B^{K^+K^-}\cos^2\Theta\right),
\nonumber\\
{\cal W}^{e^+e^-}(\cos\Theta) & = &
%\frac{3}{8}(1+\rho_{00})
%\left(1+\frac{1-3\rho_{00}}{1+\rho_{00}}\cos^2\Theta\right)
%\equiv
\frac{3}{8}(1+\rho_{00})
\left(1+B^{e^+e^-}\cos^2\Theta\right),
\label{W1}
\end{eqnarray}
where we use the normalization condition $\rho_{11}+\rho_{-1-1}+\rho_{00}=1$
and introduce the $\phi$ decay anisotropies $B$
\begin{eqnarray}
B^{K^+K^-}& = &-({1-3\rho_{00}})/({1-\rho_{00}}),\nonumber\\
B^{e^+e^-}& = &({1-3\rho_{00}})/({1+\rho_{00}}).
\label{W2}
\end{eqnarray}

\section{One-Boson Exchange Model}

Similar to previous analyses of the bremsstrahlung in $p p$ reactions
\cite{Giessen2} we use here the OBE model to parameterize the
interaction in the reaction $N N \to N N \phi$ by the set of relevant
tree level diagrams with one meson exchange line.
These reaction channels are depicted in Fig.~1. In diagram 1a
the $\phi$ is produced in an internal conversion
in a $\phi\rho\pi$ vertex, while diagram 1b
is the contribution of the $\phi$ ''bremsstrahlung''
due to a finite coupling of the $\phi$ to the nucleon current,
which might arise microscopically from the coupling of the $\phi$
to the kaon cloud around the nucleon.
Diagram 1c depicts a more exotic channel, namely
the $\phi$ shake-off reaction by the internal virtual meson flow analog
to the $\phi$ photoproduction \cite{Titov1,Henley}.
It represents the emission from the vertex.
(As alternative one could consider
the $s\bar s$ knock-out, but we assume here that the corresponding
meson-$s\bar s$-$\phi$ coupling constants are small.)
Within the OBE model the diagrams 1a - c cover basically all
reaction mechanisms.
One expects that diagram 1a
represents the dominating contribution to the OBE mechanism,
and we call this process hereafter the conventional or pure OBE
contribution.

\subsection {Emission from internal meson line}

This part of our OBE model is similar to the previously employed
model \cite{Ko_PL}.
The meson-nucleon and the $\phi\rho\pi$ interaction Lagrangians read
in obvious standard notation
\begin{eqnarray}
{\cal L}_{MNN} & = &
-i g_{\pi NN} \bar N \gamma_5 \bbox{\tau} \bbox{\pi} N
- g_{\rho NN} \left( \bar N \gamma_\mu \bbox{\tau} N \bbox{\rho}^\mu
-\frac{\kappa_\rho}{2M_N} \bar N \sigma^{\mu\nu} \bbox{\tau}N
\partial_\nu\bbox{\rho}_\mu \right),
\label{L_MNN}\\
{\cal L}_{\phi\rho\pi} & = &
g_{\phi\rho\pi} \, \epsilon^{\mu\nu\alpha\beta} \,
\partial_\mu \phi_\nu \, {\rm Tr} ( \partial_\alpha \rho_\beta \pi),
\label{L_PRP}
\end{eqnarray}
where  ${\rm Tr}(\rho \pi) = \rho^0 \pi^0 + \rho^+ \pi^- + \rho^- \pi^+$,
and bold face letters denote isovectors.
All coupling constants are dressed by monopole form factors
%\begin{eqnarray}
$F_i=(\Lambda_i^2-m_i^2)/(\Lambda_i^2-k_i^2)$,
%\label{FF}
%\end{eqnarray}
where $k_i$ is
the four-momentum of the exchanged meson.

The total invariant amplitude is the sum of 4 amplitudes
\begin{equation}
%T^r = T^r[ab;cd] + T^r[ba;dc] +  %{\rm exchanged};c\to d
%         - \xi \{ T^r[ab;dc] + T^r[ba;cd] \},
T^r_\alpha =
\sum_{m_{c,d}}
\left(\xi_\alpha^1 T^r[ab;cd] + \xi_\alpha^2 T^r[ab;dc]
+\xi_\alpha^3  T^r[ba;dc]  + \xi_\alpha^4 T^r[ba;cd]\right)
%\label{thr1}
\label{ampl1}
\end{equation}
with
$\xi^1_{pp}=\xi^3_{pp}=-\xi^2_{pp}=-\xi^4_{pp}=1$,
$\xi^1_{pn}=\xi^3_{pn}=-1$,
$\xi^2_{pn}=\xi^4_{pn}=2$.
The last two terms stem from the antisymmetrization or
from charged meson exchange for $pp$ or 
$pn$ reactions, respectively.
The first term in Eq.~(\ref{ampl1}) for the $pp$ reaction reads 
\begin{equation}
T^r_{m_cm_d;m_am_b}[ab;cd]=
K(k_{\pi}^2,k_{\rho}^2)\,
\Pi_{m_dm_b}(p_b,p_d)\,W_{m_cm_a}^r(p_a,p_c)\,
\end{equation}
with
\begin{eqnarray}
K(k_{\pi}^2,k_{\rho}^2)
& = &
\frac{g_{\pi NN} \, g_{\rho NN} \, g_{\phi\rho\pi}}
{(k^2_{\pi}-m_\pi^2)(k^2_{\rho}-m_\rho^2)}\,
\frac{\Lambda_\pi^2-m_\pi^2}{\Lambda_\pi^2-k^2_{\pi}}
\frac{\Lambda_\rho^2-m_\rho^2}{\Lambda_\rho^2-k^2_{\rho}}
\frac{{\Lambda^{\rho \, 2}_{\phi\rho\pi}}-m_\rho^2}
{{\Lambda^{\rho 2}_{\phi\rho\pi}}-k^2_{\rho}}
\frac{{\Lambda^{\pi \, 2}_{\phi\rho\pi}}-m_\pi^2}
{{\Lambda^{\pi 2}_{\phi\rho\pi}}-k^2_{\pi}}
,\nonumber\\[2mm]
\Pi_{m_dm_b}(p_b,p_d)
& = &
\bar u(p_d,m_d)\gamma_5\, u(p_b,m_b),\nonumber\\[2mm]
W_{m_cm_a}^r(p_a,p_c)
& = &
i \, \epsilon_{\mu\nu\alpha\beta} \,
k_{\rho}^\mu \, \Sigma^\nu_{m_c m_a}(k_{\rho}) \,
q_\phi^\alpha \,
\varepsilon^{*\,r,\beta},
\label{ampl_not} \nonumber \\
\Sigma^\nu_{m_cm_a}(k_{\rho})
& = &
\bar u(p_c,m_c)\left[\gamma^\nu - i \frac{\kappa_\rho}{2M_N}\sigma^{\nu\nu'}
{k_{\rho}}_{\nu'}\right]\, u(p_a,m_a).
\label{Ampl2}
\end{eqnarray}
Here $\varepsilon^{r,\beta}$ denotes the $\phi$ polarization four-vector,
and $m_a...m_d$ are the nucleon spin projections
on the quantization axis, and $k_{\rho}=p_c-p_a, \,k_{\pi}=p_b-p_d$;
$u$ stands for a Dirac bispinor.
Detailed expressions of the functions $\Pi, W$ and $\Sigma$ 
are displayed in the Appendix.

The amplitude in the $\phi$ helicity frame is obtained from the  amplitude 
in the c.m.s. by a standard Wigner rotation.

\subsection{Emission from external legs}

The amplitude of direct $\phi$ meson
emission from the nucleon legs according to Fig.~1b
is calculated similar
to the real or virtual photon
bremsstrahlung \cite{Giessen,TKB} in the few GeV region.
The $\phi NN$ vertex has the same structure
as the $\rho NN$ interaction
(cf.\ second part of Eq.~(\ref{L_MNN})), i.e., it possesses vector and tensor
parts. Following Ref.~\cite{MMSV} we use $g_{\phi NN}=-0.24$
and $\kappa_\phi=0.2$. 

The internal zig-zag lines in Fig.~1b correspond
to a suitably parameterized $NN$ interaction in the form of the Born term 
of the OBE model with effective coupling constants and cut-off parameters
and may be interpreted as effective
$\pi,\omega,\rho,\sigma$ meson exchanges.
We would like to stress
that this is an effective dynamical model
which is appropriate in the few GeV region and
which is different from the OBE model in the conventional sense.
% being reliable at much lower energies.
This model has been applied successfully
to different reactions \cite{Giessen,TKB,Giessen2} and
this encourages us to employ it for the $\phi$ production too.

The total amplitude for the process 1b consists of $4 \cdot 8$
contributions and has a similar structure as 
Eq.~(\ref{ampl1}), where $T[ab;cd]$ now reads
\begin{eqnarray}
&&{T^r[ab;cd]}
= - g_{\phi NN} 
\sum_{m=\pi,\sigma,\rho,\omega}
\left( \bar u(p_c)
\left[\Gamma^m\,\Omega_1^r+\Omega_2^r\Gamma^m\right] u(p_a)\right)
[-i\,D^m] \left(\bar u(p_d)\Gamma^m u(p_b)\right),
\nonumber\\\nonumber\\
&&\Omega_1^r=\frac{\hat p_1+M_N}{p_1^2 - M_N^2}\,
{\varepsilon^*_\mu}^r[\gamma_mu 
+i\frac{\kappa_\phi}{2M_N}\sigma^{\mu\nu}q_\nu],\,\,\,
\Omega_2^r={\varepsilon^*_\mu}^r[\gamma_\mu 
+i\frac{\kappa_\phi}{2M_N}\sigma^{\mu\nu}q_\nu]\,
\frac{\hat p_2 + M_N}{p_2^2 - M_N^2}.
\label{tbr1}
\end{eqnarray}
Here we use the notation
$p_1= p_a-q$, $p_2=p_c+q$, and
$\Gamma^m$ and $D^m$ are effective
coupling vertices and propagators of the two-body $T$ matrix, respectively,
\begin{eqnarray}
&&D^{\pi/\sigma}=\frac{i}{k^2-m_{\pi/\sigma}^2},\,\,\,\,
{D_{\mu\nu}^{\rho/\omega}}=
-i\frac{g_{\mu\nu}- k_\mu k_\nu \, m^{-2}_{\rho/\omega}}
{k^2-m_{\rho/\omega}^2},\nonumber\\ 
&&
{\Gamma^{\pi}}= -ig_{\pi NN}\gamma_5,\,\,\,\,
{\Gamma^{\sigma}}= g_{\sigma NN},\,\,\,\,
{\Gamma_\mu^{\rho/\omega}}=
- g_{\rho/\omega\,NN}\,\left(  
\gamma_\mu + i\frac{\kappa_{\rho/\omega}}{2M_N}\sigma^{\mu\nu}k_\nu \right),
\label{POBE}
\end{eqnarray}
where $k$ is the momentum of virtual meson and $g_{mNN}$ is the corresponding 
coupling constant which is dressed by the above form factors.
% in the form of Eq.~(\ref{FF}). 
The explicit  expression of the total amplitude is rather 
cumbersome and will not displayed it here. The most economic way 
to evaluate this amplitude is
a direct numerical calculation of the Feynman diagrams including
the spinor algebra. 
This method has been checked in a number 
of examples (including the OBE amplitude in Fig.~1a).

\subsection{Emission from vertex}

The main ingredient of the $\phi$ shake-off mechanism
(cf.\ Fig.~1c) is the
assumption that the constituent quark wave function of the proton contains,
in addition to the usual $uud$ component,
a configuration with an explicit $s \bar s$
contribution~\cite{Henley,Lipkin}.
A simple realization of this picture is the following wave function in the
Fock space \cite{Henley}
\begin{equation}
|p \rangle = A | [uud]^{1/2} \rangle
+
B \Bigl\{
a_0 | \bbox{[} [ uud]^{1/2} \otimes [s\bar s]^0 \bbox{]}^{1/2} \rangle
+
a_1 | \bbox{[} [ uud]^{1/2} \otimes [s\bar s]^1 \bbox{]}^{1/2} \rangle
\Bigr\},
\label{eq:wf}
\end{equation}
where $B^2$ denotes the strangeness admixture ($|A|^2+|B|^2=1$), and
$a_{0,1}^2 = \frac12$
are the fractions of the $s\bar s$ pair with spin 0 and 1.
The superscripts represent the spin of each cluster, and $\otimes$
indicates the vector addition of spins of the $uud$ and $s\bar s$ clusters
and their relative orbital angular momentum ($\ell=1$).
Details of the wave functions in the relativistic harmonic oscillator
model can be found in \cite{Titov1}.
Some estimates regarding
the strangeness content in the nucleon advocate a few
percent \cite{Ellis,newexp}.
Otherwise, calculations based on the inclusion of
extra kaonic degrees of freedom within
a perturbative treatment give more than an order of magnitude
smaller value of hidden strangeness \cite{Titov1,Henley3} which
is, however, likely to be dominated by nonperturbative effects \cite{Ioffer}.
That is the reason why we consider $\phi$ bremsstrahlung
and $\phi$ shake-off independently.

We assume here as above that the exchanged effective
mesons interact with the
$uud$ cluster as with a structureless particle and parameterize
this interaction within the effective two-body $T$ matrix.
The $s\bar s$ component is considered as spectator, that means only
the configuration with spin $S_{s\bar s}=1$ is realized.
The corresponding T matrix element
is calculated in the rest frame of the splitting  nucleon 
$(p_a=(M_N,{\bf 0}))$
and it is expressed by
\begin{equation}
{T^r_{m_cm_d;m_am_b}[ab:cd]}_{s.o.} =
\sum_{m=\pi,\rho,\omega,\sigma} \sum_{m_x}
M^{r \, N\to N\phi}_{m_x;m_a}\,
T^{(m)\,NN\to NN}_{m_c m_d;m_xm_b}(p_c,p_d;p_a',p_b)
\label{ss2}
\end{equation}
with $T^{(m)\,NN\to NN}$ as effective two-body T matrix
\begin{eqnarray}
T^{(m)\,NN\to NN}_{m_c m_d;m_xm_b}(p_c,p_d;p_a',p_b)=
\left( \bar u(p_c,m_c) \Gamma^m \,u(p_a'm_x)\right)\,[-i D^m] \,
\left(\bar u(p_d,m_d) \Gamma^m \, u(p_b,m_b)\right)
\end{eqnarray}
and with the vertices and propagators as in Eq.~(\ref{POBE}).
The transition amplitude reads
\begin{eqnarray}
M^{r \, N\to N\phi}_{m_x;m_a}
& = &-iC_M 
%C_M=-iA^*Ba_1
%\frac{V({\bf q}_l)}{ \gamma_\phi \, \gamma^{2}_c}
\sqrt{\frac32}\,
\sum
%_{j=0,1; m_j m}
\left\langle 1 m_\phi 1m\Bigl\vert j m_j \right\rangle\,
\left\langle j m_j \frac12 m_x
\Bigl\vert \frac12 m_a \right\rangle\,\hat {q}^L_m \,
d^1_{m_\phi r}(\vartheta),
\nonumber
\end{eqnarray}
with
\begin{eqnarray}
C_M & = & A^*Ba_1
\frac{V({\bf q}^L)}{ \gamma_\phi \, \gamma^{2}_c},
\label{ss4}
\end{eqnarray}
and  $p_a'=zp_a$, where $z \simeq0.5$ is the momentum fraction carried
by the $uud$ cluster,
${\bf q}^L$ denotes the $\phi$ momentum in the laboratory %(anilab.)
system, $\hat {q}^L_m$ is its projected unit vector
in the circular basis,
and $V({\bf k})$ stands for the wave function of the relative motion
normalized as
$\int  d{\bf k} \, V^2({\bf k}) /
((2\pi)^3 2 \sqrt{({\bf k}^2+M_\phi^2)}) = 1$;
$\gamma_{\phi,c}$ is the corresponding Lorentz factor
which reflects the Lorentz contraction in the relativistic
constituent model. In our calculations we use a
Gaussian distribution $V(x) = N x \exp(-x^2/2\Omega)$ with
$\sqrt{\Omega}= 2.41$ fm$^{-1}$ \cite{Titov1,Henley}. 

By making use of the  numerical values of the Clebsch-Gordan coefficients
we get the final  expression for the amplitude $M^{N\to N\phi}$
in the form
\begin{eqnarray}
M^{r N\to N\Phi}_{m_x;m_a}
=&& -i\,C_M\, \sqrt{\frac12}\sum_{m_\phi=0.\pm1}
\left(\right.\delta_{{m_\phi}0}
\left(\right.
-\hat{q}^L_z\,\delta_{m_a m_x}
+ (2m_a)\delta_{-m_am_x}\, \hat{q}^L_\perp
\frac{1}{\sqrt{2}}
\left.\right)
\nonumber\\
&&+\delta_{|{m_\phi}|1}
\left(\right.
\hat{q}^L_z\,\delta_{-m_am_x}\,\delta_{m_a m_\phi/2}
+\delta_{m_a m_x}\,\hat{q}^L_\perp
\left(\frac{m_\phi}{\sqrt{2}}-m_a\right)
\left.\right)\left.\right)d^1_{m_\phi r}(\vartheta).
\label{ss10}
\end{eqnarray}
The two body  scattering amplitude
$T^{(m)\,NN\to NN}_{m_c m_d;m_xm_b}$ for each exchange  meson is found
straightforwardly by using the functions $\Pi$ and $\Sigma$ in the Appendix.
The final amplitude
contains a sum over all exchanged mesons and consists
of 2 direct and 2 exchange amplitudes for the $pp$ reaction, and 2 direct
and 0 ($\omega, \sigma$) or 2 ($\pi, \rho$)
exchange amplitudes for the $pn$ reaction to be taken with their
proper isospin factors $\xi_{pp(pn)}^m$ as in Eq.~(\ref{ampl1}).

\section{Fixing parameters}

The parameters of the two-body $T$ matrix for $\phi$ bremsstrahlung 
(Fig.~1b) and shake-off (Fig.~1c) are taken from
Refs.~\cite{Giessen2,Giessen} where quite reasonable agreement with data
of different elastic and inelastic $NN$ reactions is found.

The coupling constant $g_{\phi\rho\pi}$
is determined by the $\phi\to\rho\pi$ decay.
The recent value $\Gamma( \phi\to\rho\pi) =$ 0.69 MeV
%\cite{PDG}
results in $g_{\phi\rho\pi}$ = 1.10 GeV$^{-1}$.
The remaining parameters of the OBE amplitude for the process
1a are taken from the
Bonn model as listed in Table B.1 (Model II) of Ref.~\cite{Bonn}.
The cut-off parameter $\Lambda^\rho_{\phi\rho\pi} = 2.9$ GeV is adjusted
by a comparison of additional calculations including the channels
discussed above and data \cite{pipnphi} for the $\pi^- p\to n\phi$
reaction. Note, that on this stage the result is not sensitive
to the specific OBE model of the $\pi p$ reaction, 
where $\Lambda^\rho_{\phi\rho\pi}$ is used as fit parameter,
that is, one can fit the data also with any other OBE model 
by using $\Lambda^\rho_{\phi\rho\pi}$ as a free parameter.
Turning to the $pp$ reaction, however, we find that 
the cut-off  $\Lambda^\pi_{\phi\rho\pi}$ represents the main
uncertainty of the model. In would be perfect to find it 
from the unpolarized cross section in this energy region and after 
that to study the polarization
observables. Lacking precise data one has to use 
some reasonable guess which make our finite energy prediction 
rather qualitative estimates than exact predictions. 
However, as we show in the next section,
some of the threshold prediction are ''parameter independent''.
In any case,
using our  model one can easily refine our finite energy prediction
when the needed data on unpolarized $\phi$ production are available and 
fix  $\Lambda^\pi_{\phi\rho\pi}$.
The value  $\Lambda^\pi_{\phi\rho\pi} =$ 0.77 GeV is provided by an
analysis of
the photoproduction of vector mesons \cite{Friman}, and we use here this
value having in mind,
however, that the symmetry of the off-shell mesons in the $\phi\rho\pi$ vertex 
would suggest $\Lambda^\pi_{\phi\rho\pi} = \Lambda^\rho_{\phi\rho\pi}$
which can be employed as another reference value. 

Note that the OBE model describes the data \cite{pipnphi}
in a limited region of
$\Delta_{\pi^- p \to n \phi} \equiv \sqrt{s}-M_N-M_\phi \le 0.15$ GeV.
At higher energies the OBE model overestimates the data.
(In this region
one could use an energy depending  suppression factor at the vertices as
in Ref.~\cite{Sibirtsev}).

Another type of uncertainty is related to the total phase 
($\eta=\exp(-i\delta)$) 
of the two-body $T$ matrix. In the effective theory  of
Refs.~\cite{Giessen,Giessen2} one fixes only the relative phase
between different exchange amplitudes and, therefore,  there are
different possible solutions. The most natural one is the assumption
that each exchange amplitude has the same phase as the pure amplitude
of the process 1a, which is known. 
In this case the two-body $T$ matrix must be real ($\delta=0$).
In that instant we expect a strong interference between the 
radiation from the internal line and the external legs, because
both amplitudes are almost real (at least in coplanar geometry), and
a small interference with the radiation from the vertex, where the amplitude
is almost imaginary. The opposite limit is based on the assumption that 
the unitary condition $SS^+=1$ and the relation $S=1-iT$ result in 
a forward scattering amplitude which is almost imaginary and negative.
The assumption that the effective $T$ matrix is responsible for 
the elastic forward scattering constrains the common phase,  
at least at small momentum transfer ($\delta=\pi/2$).
In this case one gets a strong interference between the OBE amplitude 1a
and the radiation from vertex 1c, and a small one between these and the 
radiation from the legs 1b. 
The latter  assumption is more reliable for 
high energy scattering, where the diffractive scattering 
(i.e. Pomeron exchange)
is dominant; in near-threshold $\phi$ production region it has rather 
a methodical interest. Nevertheless, for completeness we will consider 
both cases.   

\section{Threshold limit}

Let us now consider in some detail the threshold limit, where
one can neglect
terms proportional to $|{\bf q}|/M_\phi$, $|{\bf p}_{c,d}|/M_N$
in the amplitude. Spin observables in $pp$ reaction are universal
and do not depend on the specific model. So, for simplicity we
first consider the OBE amplitude for the process 1a. 

In the threshold limit the
meson propagators and form factors become constants
because they depend on the same variable
$k^2_{\rho,\pi}\to k_0^2 = - M_NM_\phi$, and using the explicit form
of the functions $\Pi,\Sigma$ and $W$ (cf. Appendix) we get
\begin{eqnarray} 
\Pi_{m_dm_b}&=&\sqrt{M_NM_\Phi}\,(2m_b)\,\cos\theta_b\,\delta_{m_dm_b},
\nonumber\\
\Sigma^x_{m_cm_a}&=&(1-\kappa_\rho)\sqrt{M_NM_\Phi}\,(2m_a)\cos\theta_a\,
\delta_{-m_am_c},
\nonumber\\
\Sigma^y_{m_cm_a}&=&i (1-\kappa_\rho)\sqrt{M_NM_\Phi}\cos\theta_a\,
\delta_{-m_am_c}\nonumber\\
W^r_{m_cm_a}&=&\frac12(1-\kappa_\rho)M_\Phi^2\sqrt{M_N(4M_N+M_\Phi)}\,
\left\{\varepsilon^{*r}_x +(2m_a)i\varepsilon^{*r}_y\right\}
\,\delta_{-m_am_c}.
\end{eqnarray}
$\Sigma^{0,z}$ is not operative here, and one can express
the amplitude in the simple form
\begin{eqnarray} 
T^r[ab;cd] =  T_0 \, U^r[ab;cd]
\nonumber
\end{eqnarray}
with
\begin{eqnarray}
T_0
& = &
(1-\kappa_\rho)\,
K(k_0^2)\,
M_NM_\phi^{5/2}(M_N + \frac 14 M_\phi)^{1/2},
\nonumber\\
K(k_{0}^2)
&=&
\frac{g_{\pi NN} \, g_{\rho NN} \, g_{\phi\rho\pi}}
{(k^2_{0}-m_\pi^2)(k^2_{0}-m_\rho^2)}\,
\frac{\Lambda_\pi^2-m_\pi^2}{\Lambda_\pi^2-k^2_{0}}
\frac{\Lambda_\rho^2-m_\rho^2}{\Lambda_\rho^2-k^2_{0}}
\frac{{\Lambda^{\rho 2}_{\phi\rho\pi}}-m_\rho^2}
{{\Lambda^{\rho 2}_{\phi\rho\pi}}-k^2_{0}}
\frac{{\Lambda^{\pi 2}_{\phi\rho\pi}}-m_\pi^2}
{{\Lambda^{\pi 2}_{\phi\rho\pi}}-k^2_{0}},
\nonumber\\[2mm]
U^r[ab;cd]
& = &
-4\,{\sqrt2}\,m_am_b
\cos\theta_b\,
\delta_{-m_a m_c}\,\delta_{m_bm_d}\,
d^1_{2m_a r}(\theta).
\label{tnr}
\end{eqnarray}
%where $d^1_{mm'}(\theta)$ is the Wigner rotation function.
The spin is transferred to the $\phi$ meson
only in the $NN\rho$ vertex by a nucleon spin-flip, and for
$T^{\pm1}(\theta=0)\neq 0$ only for $m_a=\pm\frac12$.
Summation of $|T^r|^2$ over the spin projections $m_{c,d}$ of
the outgoing particles leads to
%the following expressions for $pp$ and $pn$ collisions
\begin{eqnarray}
\sum_{m_{c,d}}|T^r_\alpha|^2 & = &
2\,T^2_0\, \Delta_{m_am_b}^\alpha \,|d^1_{2m_a \, r}(\theta)|^2,
\label{thr2}
\end{eqnarray}
where $\Delta_{m_am_b}^{pp} = 8\,\delta_{m_am_b}$,
and $\Delta_{m_am_b}^{pn} =  2\,(1+8\,\delta_{m_am_b})$.
Using these equations we get the threshold limits for the beam - target
asymmetry $C^{BT}_{zz} = 1$ for  $pp$  collisions.
The diagonal spin density
matrix elements are
$\rho_{00}^{\rm thr}=\frac12\,\sin^2\theta$ and
$\rho_{\pm1\pm1}^{\rm thr}=\frac12(1+\cos^2\theta)$.
Note that, as we mentioned above for the $pp$ reaction, this predictions 
are universal because they reflect spin and parity conservation for two 
identical fermions and do not depend on the specific mechanism. 
All of the above discussed amplitudes are satisfy the same relations.

For the $pn$ reaction this kind of predictions depend on the model 
and generally on model parameters, say, on the relative contribution 
of neutral and charged meson exchange to the two-body $T$ matrix 
in $\phi$ bremsstrahlung, shake-off
and others. However, for the OBE amplitude 1a the threshold predictions 
for the beam-target 
asymmetry and spin-density matrix are parameter independent. 
From Eq.~(\ref{thr2}) we get for the beam-target
asymmetry $C^{BT}_{zz} = 0.8$,
which means that the ratio of singlet to triplet spin states
is $\frac 19$.
The diagonal spin density matrix elements are the same as in the 
$pp$ reaction.

Using Eq.~(\ref{thr2}) we can make an important  prediction:
the ratio
of the unpolarized cross sections in $pn$ and $pp$ reactions is 
$\sigma_{np}/ \sigma_{pp}=5/2\cdot \frac12=5$,
where the factor $\frac12$ in the $pp$ cross section
represents the symmetry factor. This prediction does not depend on the OBE
model parameters.
As we shall see below, the OBE amplitude 1a dominates, 
hence fore we can expect that the ratio of the corresponding 
total cross sections would be close to the factor 5.

\section{Results}

Our results for the total cross section of the $pp$ reaction are
shown in Fig.~2 by the solid lines.
As we mentioned above, the comparison of the momenta at 
the $\pi \rho \phi$ vertex
in the $NN \to NN \phi$ reaction with the
$\pi N\to N\phi$ reaction shows a limited reliability of
the pure OBE model to the region of
$\Delta_s \equiv \sqrt{s}-2M_N - M_\phi \le 0.15$ GeV.
One can easily extend the model to a wider
energy region, say $\Delta_s \sim 1$ GeV,
by reducing the cross section with the effective cut-off factor
$\propto \exp (-0.8\,\Delta_s)$ resulting from the energy depending coupling
strengths in Ref.~\cite{Giessen2}.
The resulting cross section is shown by the dashed lines.

Triangles in Fig.~2 show our prediction for the $\phi$ bremsstrahlung
contribution 1b. Open and full circles show the predictions
for the shake-off channel 1c with
a strangeness probabilities $B^2 = 0.01$ and 0.05, respectively.
The dot-dashed lines show the contributions from the pure OBE channel 1a.
The $\phi$ bremsstrahlung and shake-off contributions have
the same order of magnitude. The shake-off channel in this reaction is
reduced strongly by the form factor $V$ in Eq.~(\ref{ss4}).
The solid lines show the coherent sum of all
amplitudes with $B^2=0.05$. The left (right) panel in Fig.~2 displays
the result for a
purely real (imaginary) phase of the two-body $T$ matrix. The
interference between the pure OBE amplitude 1a and the 
shake-off amplitude 1c in the first case
is rather small and therefore the difference between the results 
of the shake off with $a_1=\pm1$ is not seen. 
But it may be seen, in principle,
in second case, where the positive value of $a_1$ is marked by crosses.
In spite of the difference
between the separate contributions of the conventional OBE channel 1a and 
the exotic $\phi$ production channels, which
is more than an order of magnitude
at $\Delta_s \simeq 0.1$ GeV for the above given parameters
and $g_{\phi NN} \simeq -0.24$ and $B^2 \le 0.05$, the interference
may amount up to 50\% of the total cross section. 

The cross section for the $pn$ reaction is greater by a factor
$\simeq 5$ as given already in our threshold-near prediction.

The interference between the conventional OBE process 1a and the 
other exotic $\phi$ channels 1b,c is 
much stronger in polarization observables.
Fig.~3  shows our finite-energy predictions for the beam - target 
asymmetry Eq.~(1)
as a function of the $\phi$ production angle $\theta$ at fixed recoil
nucleon angles $\theta_{p'},\varphi_{p'} = \pi$
at $\Delta_s = 0.1$ GeV and
$|{\bf q}| = \frac23 \lambda(s,M_\phi^2,4M_N^2)/(2 \sqrt{s})$
for the $pp$ reaction for separate channels
($\lambda$ is the usual triangle function).
The contribution from the
OBE channel 1a is depicted by the dot-dashed line, while the
$\phi$ bremsstrahlung contribution 1b is marked by triangles, and
the $\phi$ shake-off process 1c is marked by dots.
The threshold prediction is the dashed line.

One can see that the asymmetry for the conventional OBE channel 1a and
the $\phi$ bremsstrahlung channel 1b are qualitatively very similar.
But in the case of a real phase of the two-body $T$ matrix there is a
strong ''destructive'' interference in the asymmetry which forces
a significant decrease of the total asymmetry. This is shown in 
Fig.~4 (left panel) where we display the beam - target asymmetry
as a coherent sum of all channels.
Solid, dotted and dash-dotted lines correspond to
$B^2 =$ 0., 0.01 and 0.05, respectively. The contribution from the shake-off
process is rather small. 
The strong decrease of the total asymmetry may be considered 
even as interesting quantitative manifestation of the 
$\phi$ bremsstrahlung process 1b.
If one would use instead of $g_{\phi NN} = - 0.24$ the same positive
quantity, then the asymmetry results in much larger values 
0.4 $\cdots$ 0.6; the corresponding result is marked by crosses.

The right panel of Fig.~4 illustrates the asymmetry for a purely imaginary
two-body $T$ matrix. The crosses mark a positive value of 
$a_1$ in Eq.~(\ref{eq:wf}).
The sign of $a_1$ is still unknown. The area between marked and 
unmarked dot-dashed lines represents 
the theoretical uncertainty of the predictions for $B^2=0...0.05$.
For positive values of $a_1$  the asymmetry
differs noticeably from the pure OBE process 1a prediction
and may reach even negative values.

The influence of this  uncertainty  is smaller for the spin density matrix
Eq.~(\ref{rhom}). Fig.~5 (left panel, same notation as in Fig.~3) depicts
$\rho_{00}$ separately for each
channel at $\Delta_s=0.1$ GeV.
The anisotropy  of the angular distribution
in 2-body decays $\phi\to a^+a^-$ is given by Eqs.~(\ref{W1},\ref{W2}).
One can see
in Fig.~5 (right panel, same notation as in right panel of Fig.~4)
a strong modification of the anisotropy
for $\phi \to e^+ e^-$ decays by the  shake-off mechanism at small
angles $\theta$. This effect is smaller for $K^+K^-$ decays.
%and does not depend strongly on the sign of the $s\bar s$ amplitude $a_1$.
The dotted and dot-dashed lines marked by diamonds correspond 
to calculations with real two-body $T$ matrix and $B^2$=0.01 and 0.05,
respectively. 

In $pn$ reactions all these polarization effects exist too,
but they are not so strong as in the $pp$ reactions.
That is because the OBE channel 1a here is a factor of 5 greater,
while the contribution of the ''exotic'' channels remains
on the same level as in the $pp$ case.

Finally, we would like to mention that 
the choice $\Lambda^\pi_{\phi\rho\pi} =$ 2.9 GeV
(instead of 0.77 GeV) increases the OBE contribution 1a by factor 
4 $\cdots$ 6, while the other ''exotic''contributions remain on the same level.
This results in a decrease of their relative role in polarization 
observables which then become much closer to the pure OBE channel 1a
predictions.
  
\section{Summary}

In summary we calculate within an extended OBE model
with ''exotic'' $\phi$ production channels the cross section and
polarization observables for the reaction $NN \to NN \phi$.
We predict the threshold behavior of cross sections for
polarized and unpolarized reactions.
Above the threshold
the polarization observables differ strongly
from the threshold predictions and
behave differently for each channel.
This means that  measurements of the
beam - target asymmetry and anisotropy of the $\phi$ decays
may reveal the presence of hidden strangeness and help disentangling
various reaction mechanisms.\\

{\bf Acknowledgments:}
Useful discussions with C.M. Ko, U. Mosel, A.A. Sibirtsev
and members of the DISTO collaboration are gratefully acknowledged.
One of the authors (A.I.T.) acknowledges the warm hospitality of the nuclear
theory group in the Research Center Rossendorf.
This work is supported by the BMBF grant 06DR829/1 and the
Heisenberg-Landau program.

%%%%%%%%%%%%%%%%%%%%%  Appendix  %%%%%%%%%%%%%%%%%%%%%%%%%%

\appendix

\section{Functions \bbox{$\Pi,\Sigma$} and \bbox{$ W$} for the OBE amplitude}

\noindent
\underline{$\Pi_{m_dm_b}(p_b,p_d)$:}\\
Using the definition of the Dirac spinor
\begin{equation}
u(p) = \sqrt{E_p+M} \left( \begin{array}{c} 1 \\
\alpha \bbox{\sigma} \cdot \bbox{n} \end{array} \right)\chi(m_p),
\quad
\alpha \equiv \sqrt{\frac{E_p-M_N}{E_p+M_N} },
\end{equation}
and the identity
\begin{eqnarray}
\chi^+_f\,({\bbox{\sigma}}\cdot{\bf A})\,\chi_i=
2m_iA_z\delta_{m_im_f}+(A_x+ 2im_iA_y)\delta_{m_i-m_f},
\end{eqnarray}
we find
\begin{eqnarray}
\Pi_{m_dm_b}& = &
C_{bd}  \left[ \right.
2m_d(\alpha_b\cos\theta_b - \alpha_d\cos\theta_d)\,\delta_{m_bm_d}
\nonumber\\
& - &\alpha_d\sin\theta_d\exp(-2im_d\varphi_d)\,\delta_{m_b-m_d}
\left. \right],
\label{pi}
\end{eqnarray}
where $C_{bd}=\sqrt{(E_b+M_N)(E_d+M_N)}$.
When the incoming particle moves along the $z$ axis, 
$\cos\theta_b=\pm 1$.\\

\noindent
\underline{$\Sigma^\nu_{m_cm_a}(k_\rho)$:}\\
Calculating $\Sigma^\nu (k_\rho)$, where $k_\rho=p_c-p_a$,
we use Gordon's identity
\begin{eqnarray}
\bar u(p_c)\left[\gamma^\nu\right] \, u(p_a) 
= \frac{1}{2M}
\bar u(p_c)\left[(p_a+p_c)^\nu +  i \sigma^{\nu\nu'}
{k_\rho}_{\nu'}\right]\, u(p_a).
\end{eqnarray}
which allows to express $\Sigma^\nu$ as the sum
\begin{eqnarray}
&&\Sigma_{m_cm_a}^\nu = \Sigma_{1\,m_cm_a,}^\nu + \Sigma_{2\,m_cm_a}^\nu,
\nonumber\\
&&\Sigma_{1\,m_cm_a}^\nu =
\frac{\kappa_\rho}{2M_N}P_{ac}^\nu\bar u(p_c)\, u(p_a),
\nonumber\\
&&\Sigma_{2\,m_cm_a}^\nu =
(1-\kappa_\rho)\,\bar u(p_c)\gamma^\nu \, u(p_a)
\label{Sigma_12}
\end{eqnarray}
with $P_{ac}=p_a+p_c$.
Direct evaluation of $\Sigma_1$ gives
\begin{eqnarray}
\Sigma_{1\,m_cm_a}^\nu 
& = &
\frac{\kappa_\rho}{2M_N}P_{ac}^\nu C_{ac}
[(1-\alpha_a\alpha_c\cos\theta_a\cos\theta_c)\,\delta_{m_am_c}
\nonumber\\
& - &
2m_a\,\alpha_a\alpha_c\cos\theta_a\sin\theta_c\,\exp(-2im_c\varphi_c)\,
\delta_{m_a-m_c}].
\end{eqnarray}
Explicit expressions for  the components of $\Sigma_{2}^\nu$ read
\begin{eqnarray}
\Sigma_{2\,ac}^0 =
(1-\kappa_\rho)\, C_{ac}
& [ &
(1+\alpha_a\alpha_c\cos\theta_a\cos\theta_c)\,\delta_{m_am_c}
\nonumber\\
& + &
2m_a\,\alpha_a\alpha_c\cos\theta_a\sin\theta_c\,\exp(-2im_c\varphi_c)\,
\delta_{m_a-m_c}],
\label{sigma20}\\
\Sigma_{2\,ac}^x 
(1-\kappa_\rho)\, C_{ac}
& [ &
\alpha_c \sin\theta_c \exp(-2im_c\varphi_c)\,\delta_{m_am_c}
\nonumber\\
& + &
2m_a\,(\alpha_a \cos\theta_a-\alpha_c \cos\theta_c)\,\delta_{m_a-m_c}],
\label{sigma2x}\\
\Sigma_{2\,ac}^y =i\,
(1-\kappa_\rho)\, C_{ac}
& [ &
2m_a\,\alpha_c\sin\theta_c\,\exp(2im_c\varphi_c)\,\delta_{m_am_c}
\nonumber\\
& + & (\alpha_a \cos\theta_a-\alpha_c \cos\theta_c)\,\delta_{m_a-m_c}],
\label{sigma2y}\\
\Sigma_{2\,ac}^z =
(1-\kappa_\rho)\, C_{ac}
& [ &
(\alpha_a\cos\theta_a + \alpha_c\cos\theta_c)\,\delta_{m_am_c}
\nonumber\\
& + & 
2m_a\,\alpha_c \sin\theta_c\,\exp(-2im_c\varphi_c)\,\delta_{m_a-m_c}].
\label{sigma2z}
\end{eqnarray}

\noindent
\underline {$W_{ac}^r(p_a,p_c)$:}\\
For calculating $W^r$ we use the identity
\begin{eqnarray}
\epsilon_{\mu\nu\alpha\beta}
k^\mu_\rho\,\Sigma^\nu q_\phi^\alpha \varepsilon^{*\,r,\beta}
& = &
k_{\rho 0}\, \bbox{q} \cdot (\bbox{\Sigma} \times \bbox{\varepsilon}^{*r})
 - q_0 \bbox{k_\rho} \cdot (\bbox{\Sigma} \times
\bbox{\varepsilon}^{*r})
\nonumber\\
& + &\Sigma_0 \,\bbox{\varepsilon}^{*r} \cdot (\bbox{k_\rho} \times \bbox{q})
 - \varepsilon_0^{*r}\, \bbox{\Sigma} \cdot (\bbox{k_\rho} \times \bbox{q}),
\label{w1}
\end{eqnarray}
which allows to express amplitude $W^r$ in compact form as
\begin{eqnarray}
W^r= i\,\left(\bbox{A}\cdot \bbox{C}^r  +\bbox{B}\cdot \bbox{D}^r\right),
\label{w2}
\end{eqnarray}
where
\begin{eqnarray}
& &\bbox{A} = k_{\rho 0} \bbox{q} - q_0 \bbox{k_\rho},
\quad 
\bbox{B} = \bbox{k_\rho} \times \bbox{q},\nonumber\\
&&\bbox{C}^r = \bbox{\Sigma} \times \bbox{\varepsilon}^{*r}
\quad
\bbox{D}^r = \Sigma_0\, \bbox{\varepsilon}^{*r}
-\varepsilon_0^{*r}\,\bbox{\Sigma}.
\end{eqnarray}
(for simplicity we skip here the arguments 
$p_a,\,p_c,...$ in $W^r$ and $\Sigma^r$).
The independent $\phi$ meson polarization vectors
$\varepsilon^r =(\varepsilon^r_x,\,\varepsilon^r_y ,\,
\varepsilon^r_z ,\,\varepsilon^r_0 )$ with $r=1,2,3$, read
\begin{eqnarray}
&&\varepsilon^1 =(\cos\theta,\, 0 \, -\sin\theta,\, 0),
\nonumber\\
&&\varepsilon^2 =(0 ,\, 1, \, 0,\, 0),
\nonumber\\
&&\varepsilon^3 =(\gamma_\phi\sin\theta,\, 0,\,
\gamma_\phi\cos\theta,\,\gamma_\phi v_\phi),
\end{eqnarray}
where $\gamma_\phi=E_\phi/M_\phi,\, v_\phi= |{\bf q}|/E_\phi$.
Using the polarization vectors in the helicity  basis
\begin{eqnarray}
\varepsilon^{*0}=\varepsilon^3,
\quad
\varepsilon^{* \pm 1}=\mp\frac{1}{\sqrt{2}}(\varepsilon^1\mp i \varepsilon^2 )
\end{eqnarray}
we get the amplitude in the $\phi$ helicity basis as
\begin{eqnarray}
W^0=W^3,
\quad
 W^{\pm 1}= \mp\frac{1}{\sqrt{2}}(W^1 \mp i W^2),
\end{eqnarray}
which may be directly used in numerical calculations.

\newpage

\begin{figure}[h]

\vspace*{3cm}

\centering
\includegraphics[width=17cm]{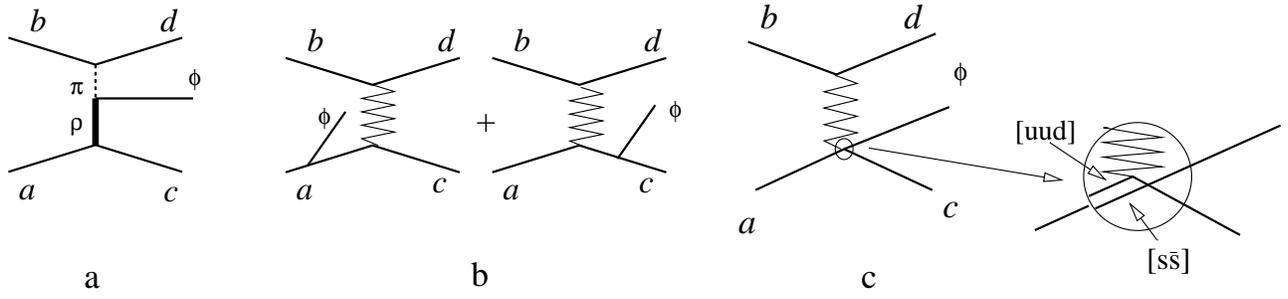}
%\begin{figure}[h]
~\\[9mm]
\caption{
Diagrammatic representation of processes of the $\phi$ production:
emission from
(a) the internal meson line,
(b) the external nucleon legs, and
(c) the exchanged meson - nucleon vertex.
Exchange diagrams are not displayed.
}
\label{fig1}
\end{figure}

\newpage

\vspace*{3cm}

\centering
\includegraphics[width=7cm]{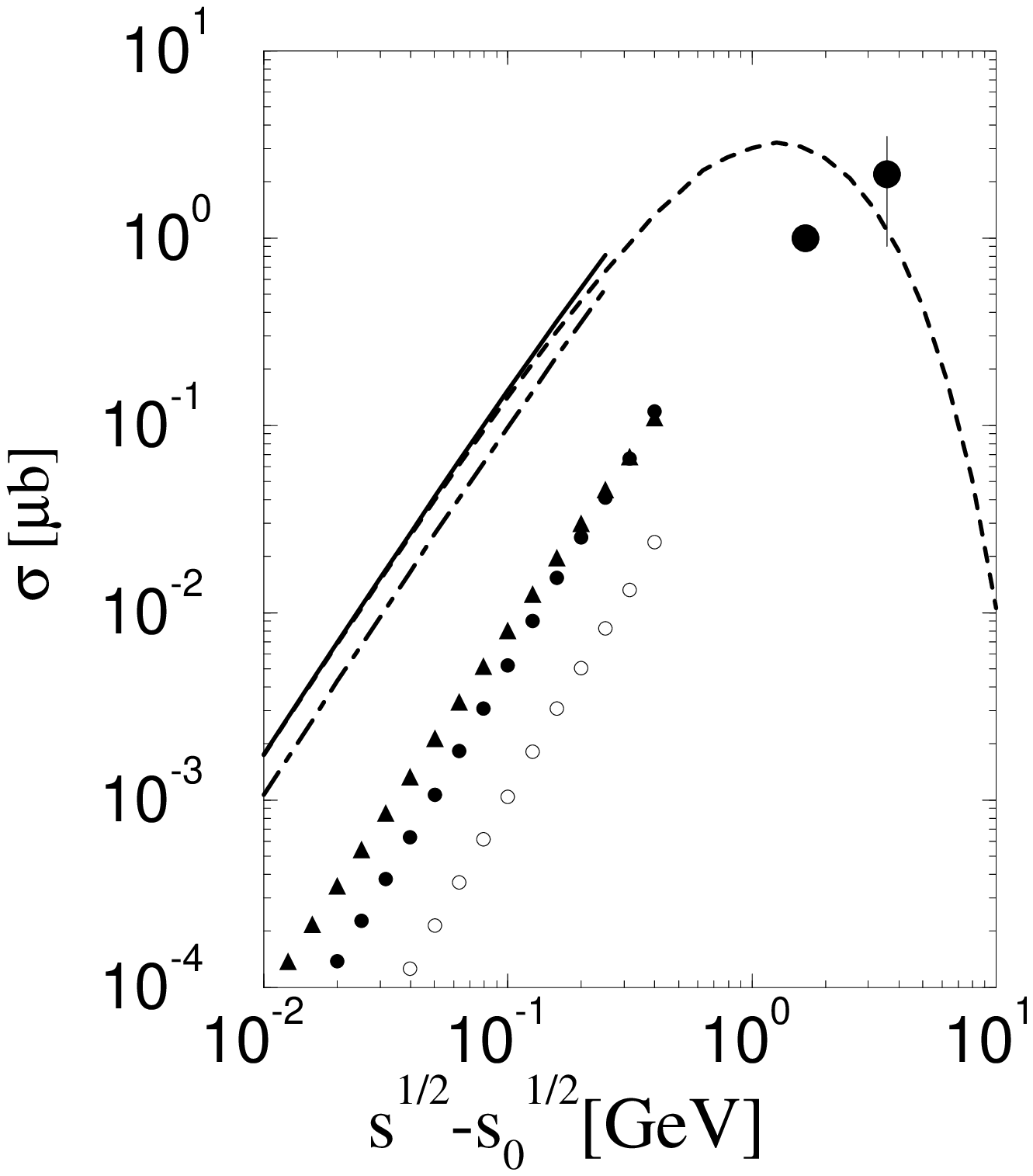} \qquad
\includegraphics[width=7cm]{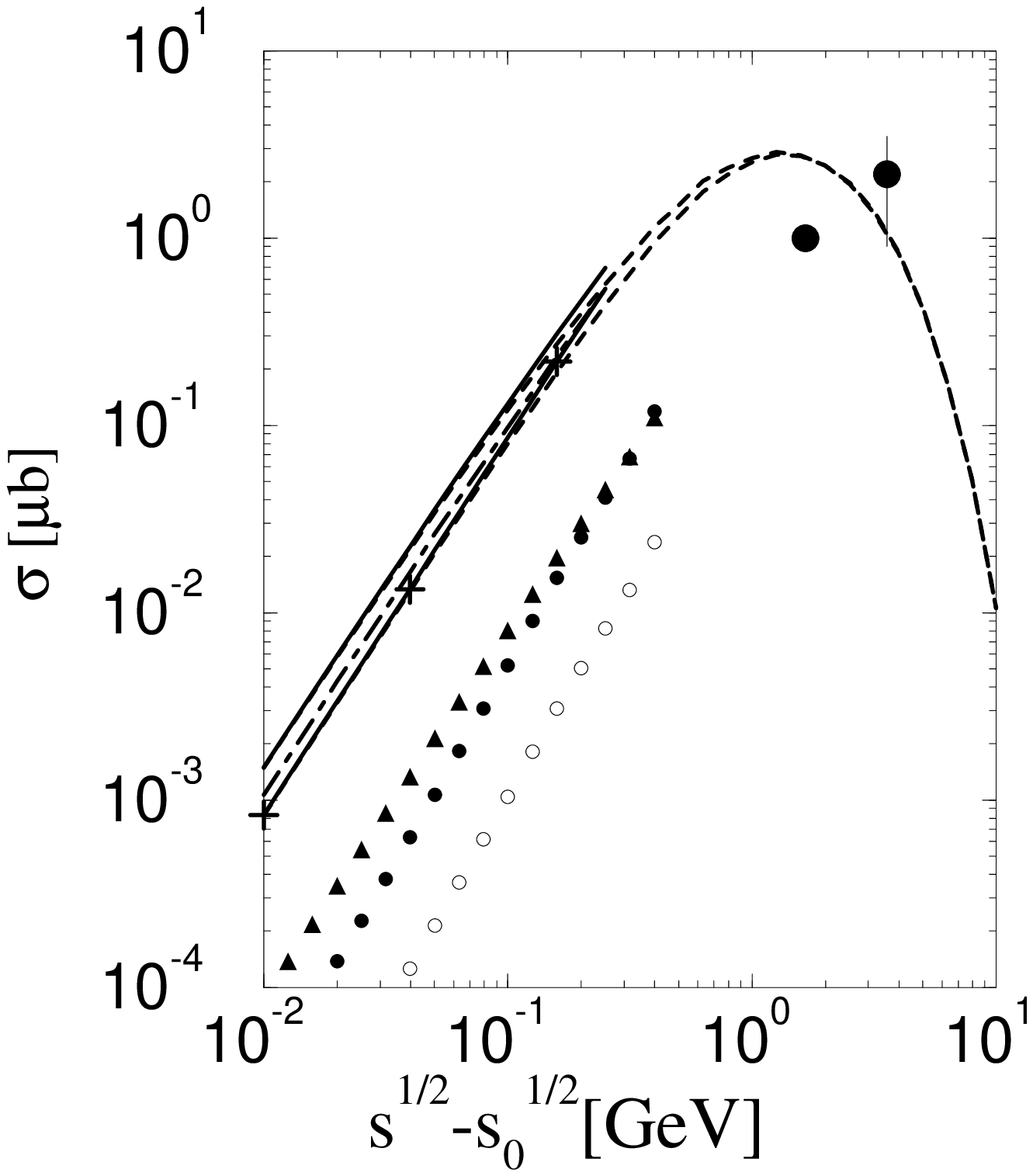}
\begin{figure}[h]
~\\[3mm]
\caption{
The total cross section of the $pp \to pp \phi$ reaction as a function
of the excess energy. Data (fat dots) from  \protect\cite{Baldi}.
The individual contributions are shown separately: 
pure OBE process 1a (dot-dashed lines),
$\phi$ bremsstrahlung 1b (triangles), 
$\phi$ shake-off (open and filled circles for $B^2 =$ 1 and 5\%).
Solids lines depict the coherent sum of all amplitudes (with $B^2$=0.05),
while dashed lines represent these sums with additional energy depending
effective cut-off described in text.
Left panel:  real two-body $T$ (the difference
between positive and negative values $a_1$ is not seen on this scale);  
right panel: purely imaginary two-body $T$ matrix
(crosses mark the coherent sum with positive value of $a_1$).
}
\label{fig2}
\end{figure}

\newpage

\vspace*{3cm}

\centering
\includegraphics[width=7cm]{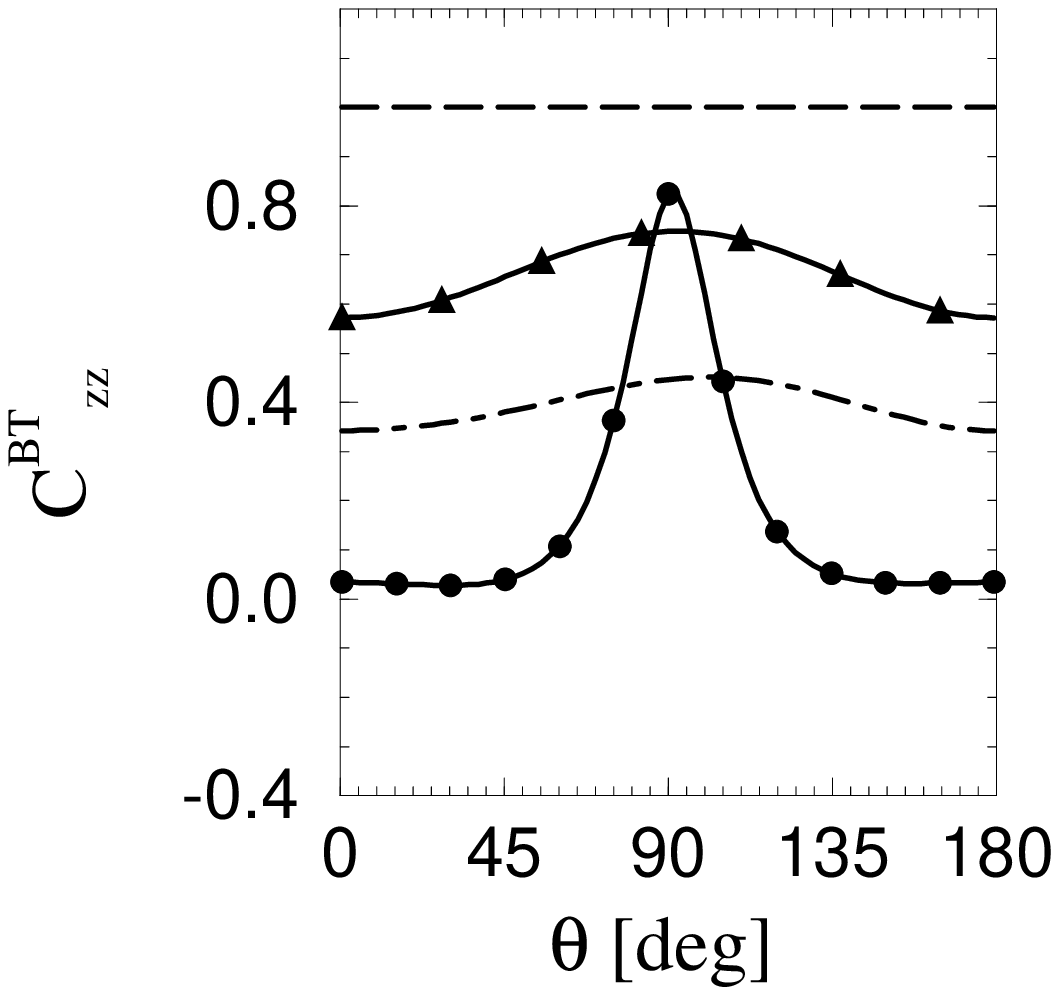}
\begin{figure}[h]
~\\[3mm]
\caption{
The finite - energy beam-target asymmetry for the $pp \to pp \phi$
reaction as a function of the c.m.s. 
$\phi$ production angle $\theta$ for different channels. 
The contribution from the
OBE channel 1a is the dot-dashed line,
the $\phi$ bremsstrahlung channel 1b is marked by triangles, and
the $\phi$ shake-off 1c is marked by dots.
The threshold prediction is depicted by the long-dashed line.
}
\label{fig3}
\end{figure}

\newpage

\vspace*{3cm}

\centering
\includegraphics[width=7cm]{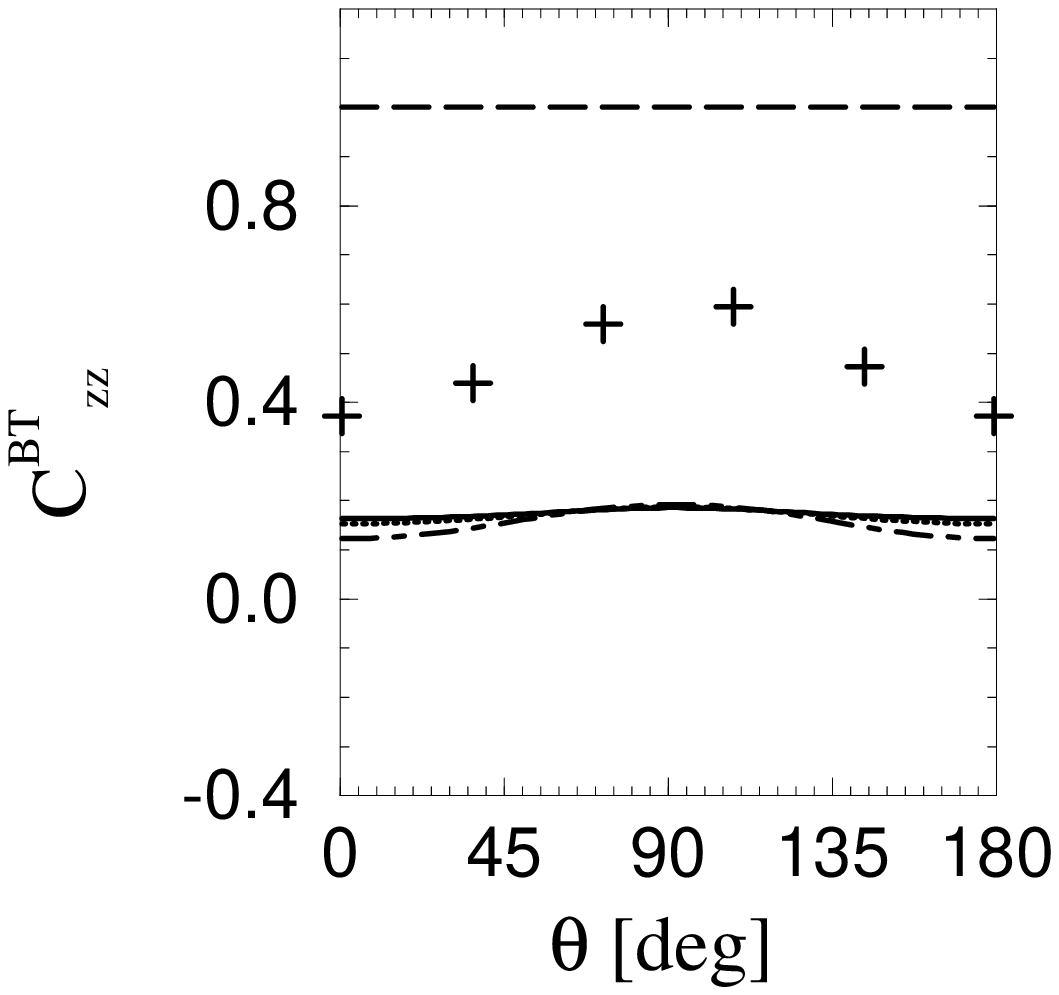} \qquad
\includegraphics[width=7cm]{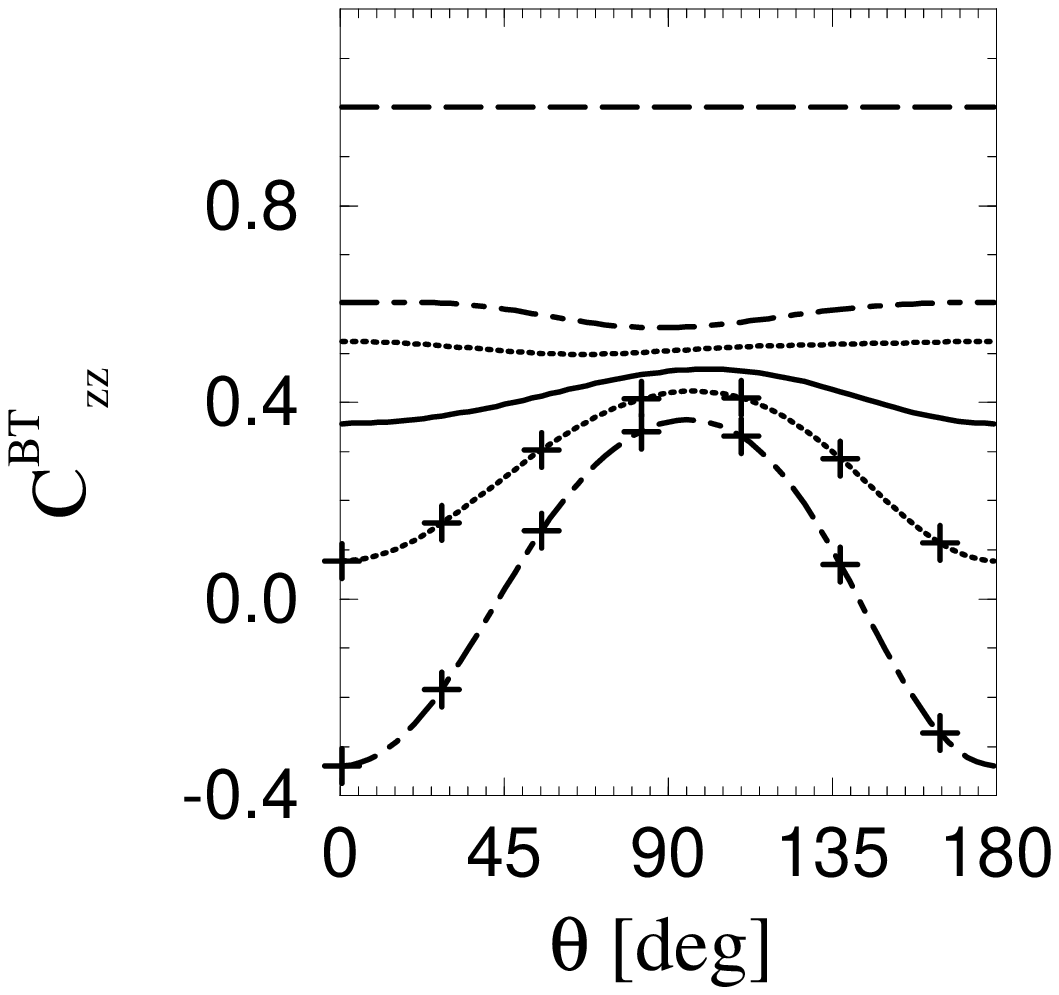}
\begin{figure}[h]
~\\[3mm]
\caption{
The beam-target asymmetry as a coherent sum of all channels.
Solid, dotted and dot-dashed lines correspond to
$B^2$=0., 0.01 and 0.05, respectively.
Left  panel: real two-body $T$ matrix (the difference between positive 
and negative $a_1$ values is not visible on this scale;
crosses indicate the result for $g_{\phi NN} = + 0.24$);
right panel: imaginary two-body $T$ matrix 
(a positive value of $a_1$ is  marked by crosses).
The threshold prediction is depicted by the long-dashed line.
}
\label{fig4}
\end{figure}

\newpage

\vspace*{3cm}

\includegraphics[width=7cm]{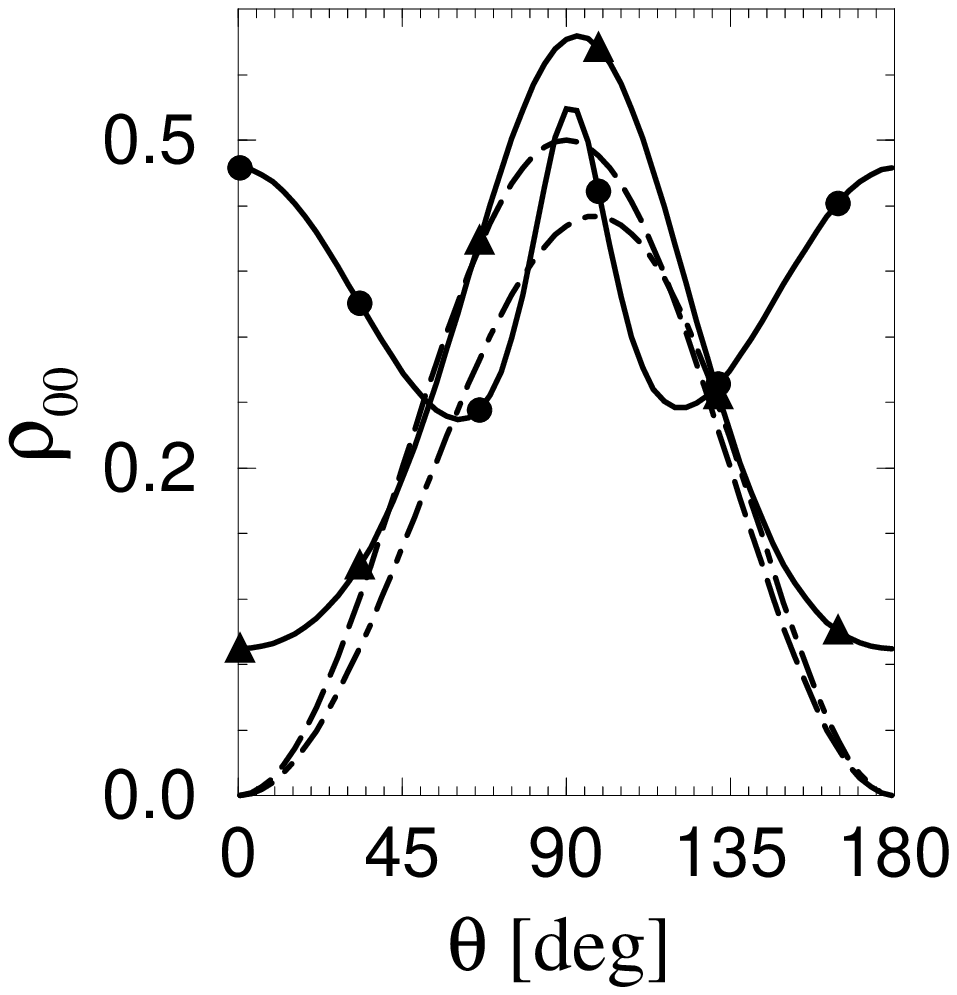} \qquad
\includegraphics[width=7cm]{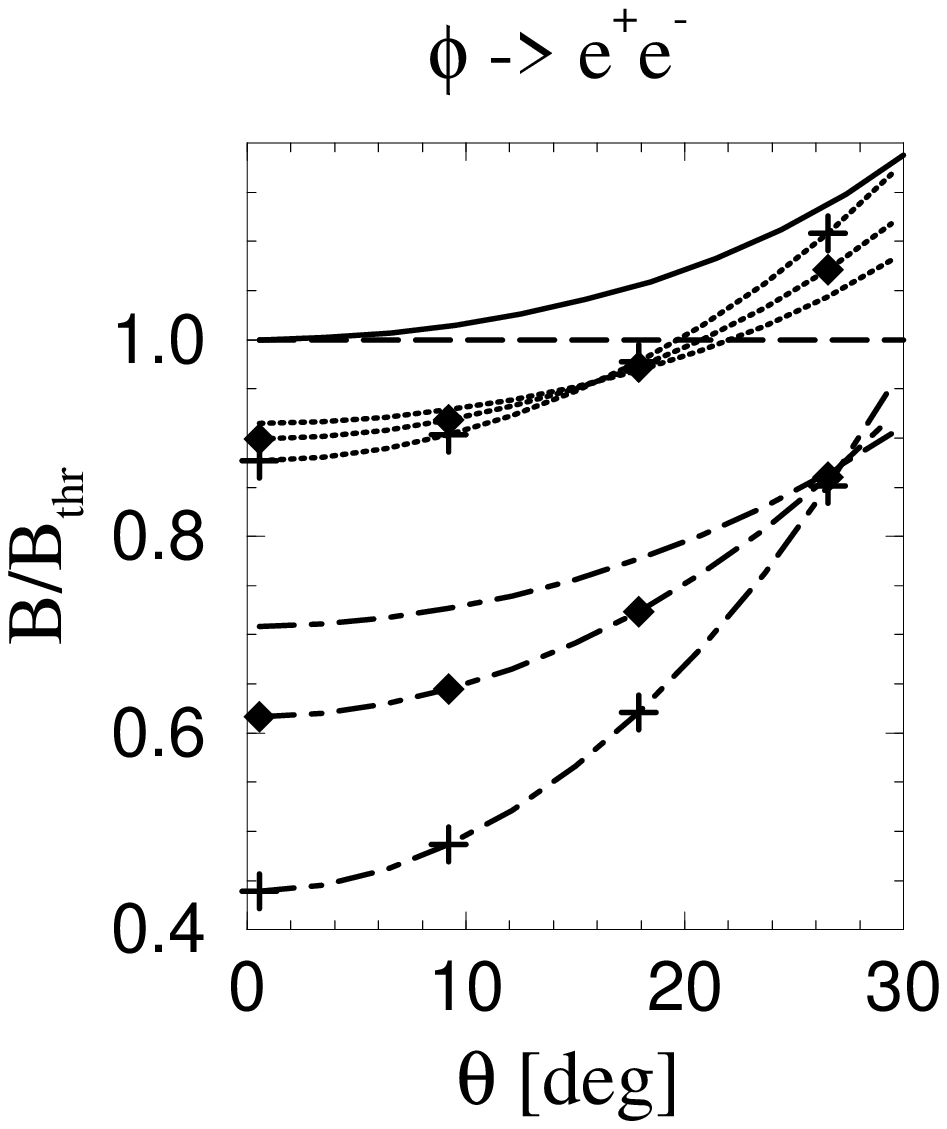}
\begin{figure}[h]
~\\[3mm]
\caption{
The spin density matrix element $\rho_{00}$ separately
for different channels (left panel;
notation as in Fig.~\ref{fig3}) and
the asymmetry of $\phi\to e^+e^-$  decays
normalized to the threshold value (right panel;
for an imaginary two body $T$ matrix; notation
as in right panel of Fig.~\ref{fig4};
the results for a real phase are marked by diamonds).
}
\label{fig5}
\end{figure}


\newpage

\begin{thebibliography}{99}
\bibitem{Sibirtsev}
A.A. Sibirtsev, Nucl. Phys. A {\bf 604}, 455 (1996).
\bibitem{Ko_PL}
W.S. Chung, G.Q. Li, C.M. Ko, Phys. Lett. B {\bf 401}, 1 (1997);
%\bibitem{Ko} W.S. Chung, G.Q. Li, C.M. Ko,
Nucl. Phys. A {\bf 625}, 347 (1997).
\bibitem{Weise}
F. Klingl, T. Wass, W. Weise, Phys. Lett. B {\bf 431}, 254 (1998). 
\bibitem{Kolomeitsev}
E.E. Kolomeitsev, D.N. Voskresensky,
B. K\"ampfer, Nucl. Phys. A {\bf 588}, 889 (1995).
\bibitem{Grosse}
R. Barth et al. (KAOS collaboration), Phys. Rev. Lett. {\bf 78}, 4007 (1997).
\bibitem{Kotte}
N. Herrmann (FOPI collaboration), Nucl. Phys. A {\bf 610}, 49c (1996).
\bibitem{HADES}
J. Friese et al. (HADES collaboration), GSI report 97-1, p. 193 (1997)
\bibitem{Brenschede}
F. Balestra et al. (DISTO collaboration), Phys. Rev. Lett. in print
\bibitem{Sapozhnikov}
M.G. Sapozhnikov et al., COSY Letter of Intent \# 35 (1995);\\
C. Wilkin, internal report to the COSY ZDF group (unpublished).
%\bibitem{OZI}
%S. Okubo, Phys. Lett. B 5 (1963) 165;
%G. Zweig, CERN report No. 8419/TH 412 (1964);\\
%I. Iizuka, Prog. Theor. Phys. Suppl. 37/38 (1966) 21
\bibitem{Ellis}
J. Ellis, M. Karliner, D.E. Kharzeev, M.G. Sapozhnikov,
Phys. Lett. B {\bf 353}, 319 (1995).
%,\\
%J. Ellis, E. Gabathuler, M. Karliner, Phys. Lett. B 217 (1989) 173
\bibitem{LZL93}
M.P. Locher, Y. Lu, Z. Phys. A {\bf 351}, 83 (1995);\\
%Y. Lu, B. S. Zou, M. P. Locher, Z. Phys. A 345, 207 (1993);\\
%H. J. Lipkin, B.-S. Zou, Phys. Rev. D 53, 6693 (1996);\\
%V. E. Markushin
% Nucl. Phys. B (Proc. Suppl.) A 56, 303 (1997).
%\bibitem{BL94}
D. Buzatu, F. M. Lev,  Phys. Lett. B {\bf 329}, 143 (1994).
\bibitem{sigmaterm}
%J.F. Donoghue, C.R. Nappi, Phys. Lett. B 168 (1986) 105,\\
J. Gasser, H. Leutwyler, M.E. Sainio, Phys. Lett. B {\bf 253}, 252 (1991).
\bibitem{newexp}
D.B. Kaplan, A.V. Manohar, Nucl. Phys. B {\bf 310} 527 (1988);\\
R.D. McKeown, Phys. Lett. B {\bf 219}, 140 (1989);\\
E.M. Henley, G. Krein, S.J. Pollock, A.G. Williams,
Phys. Lett. B {\bf 269}, 31 (1991).
%D.B. Kaplan, Phys. Lett. B 275 (1992) 137
\bibitem{Titov1}
A.I. Titov, Y. Oh, S.N. Yang, Phys. Rev. Lett. {\bf 79}, 1634 (1997);
%\bibitem{Titov2}
%A.I. Titov, S.N. Yang, Y. Oh,
Nucl. Phys. A {\bf 618}, 259 (1997).
\bibitem{Rekalo}
M.P. Rekalo, J. Arvieux, E. Tomasi-Gustafsson, 
Z. Phys. A {\bf 357}, 133 (1997).
\bibitem{Giessen2}
A. Engel, R. Shyam, U. Mosel, A.K. Dutt-Mazumder,
Nucl. Phys. A {\bf 603}, 387 (1996);\\
%\bibitem{ourbrem}
V. Shklyar, B. K\"ampfer, B.L. Reznik, A.I. Titov,
Nucl. Phys. A {\bf 628}, 255  (1998).
\bibitem{Henley}
E.M. Henley, G. Krein, A.G. Williams, Phys. Lett. B {\bf 281}, 178 (1992).
\bibitem{Giessen}
M. Sch\"afer, H.C. D\"onges, A. Engel, U. Mosel,
Nucl. Phys. A {\bf 575}, 429 (1994).
\bibitem{TKB}
A.I. Titov, B. K\"ampfer, E. Bratkovskaya, Phys. Rev. C {\bf 51}, 227 (1995).
\bibitem{MMSV}
U.-G. Meissner, V. Mull, J. Speth, J.W. Van Orden,
Phys. Lett. B {\bf 408}, 381 (1997).
\bibitem{Lipkin}
R.L. Jaffe, H. Lipkin, Phys. Lett. B {\bf 266}, 458 (1991);\\
%\bibitem{Keppler}
J. Keppler, H.M. Hofmann, Phys. Rev. D {\bf 51}, 3936 (1995).
\bibitem{Henley3}
W. Koepf, E.M. Henley, S.J. Pollock, Phys. Lett. B {\bf 288}, 11 (1992).
\bibitem{Ioffer}
B.L. Ioffe, M. Karliner, Phys. Lett. B {\bf 247}, 387 (1990).
%\bibitem{PDG}
%R.M. Barnett et al. (Particle Data Group), cf. http://pdg.lbl.gov
\bibitem{Bonn}
R. Machleidt, Adv. Nucl. Phys. {\bf 19}, 189 (1989).
\bibitem{pipnphi}
V. Flaminio, CERN preprint CERN-HERA 01/84 (1984).
\bibitem{Friman}
B. Friman, M. Soyer, Nucl. Phys. A {\bf 600}, 477 (1996).
\bibitem{Baldi}
R. Baldi {\it et al.}, Phys. Lett. B {\bf 68}, 381 (1977);
V. Blobel {\it et al.}, Phys. Lett. B {\bf 59}, 88 (1975).
\end{thebibliography}
\end{document}